\documentclass[12pt]{article}
\usepackage{amsmath,amssymb,slashed}
\usepackage[english]{babel}

\csname@addtoreset\endcsname{equation}{section}

\newcommand{\tamphys}{\it George and Cynthia Woods Mitchell  Institute
for Fundamental Physics and Astronomy,\\
Texas A\&M University, College Station, TX 77843-4242, USA}

\def\nn{\nonumber}
\newcommand{\eq}[1]{(\ref{#1})}
\newcommand{\w}[1]{\\[0.#1cm]}
\def\be{\begin{equation}}
\def\ee{\end{equation}}
\def\bea{\begin{eqnarray}}
\def\eea{\end{eqnarray}}
\def\ba{\begin{array}}
\def\ea{\end{array}}

\newcommand{\boxedeq}[1]{
\begin{equation}
\fbox{
\rule[0.7cm]{0pt}{0pt}
$#1$
\rule[-0.45cm]{0pt}{0pt}
}
\end{equation}
}

\begin{document}

\begin{flushright}
\hfill{ \ MIFPA-12-36\ \ \ \ }
\end{flushright}

\vspace{25pt}
\begin{center}

{\LARGE {\bf Supersymmetric Proca-Yang-Mills System}}

\vspace{50pt}

{\bf Ergin Sezgin and Linus Wulff}

\vspace{40pt}

{\tamphys}

\vspace{80pt}

\underline{ABSTRACT}
\end{center}

We determine the off-shell $N=1$ supersymmetry transformation rules for a tensor-Yang-Mills system in which  the tensor field transforms in a nontrivial representation of the Yang-Mills group, and there is an additional vector multiplet in the same representation. We then show that this system describes a massive tensor multiplet coupled to Yang-Mills in which the additional vector multiplet is eaten by the tensor field. Next, we construct an action which consists of four separately off-shell supersymmetric pieces and find that integrating out the auxiliary fields gives rise to an infinite number of higher order couplings of the scalar field belonging to the tensor multiplet. We describe the dualization of the massive tensor to a massive vector, thereby ending up with an on-shell supersymmetric  nonabelian massive vector multiplet coupled to an off-shell Yang-Mills multiplet which we refer to as the supersymmetric Proca-Yang-Mills system. This description of massive vectors is contrasted with the standard one obtained through the Higgs mechanism.

\vspace{15pt}

\thispagestyle{empty}



\newpage
\setcounter{page}{1}
\tableofcontents

\section{Introduction}

In this paper, we construct $N=1, D=4$ off-shell supersymmetric couplings of $n_V$ vector multiplets and $n_T$ tensor multiplets in such a way that the combined system is invariant under a semi-simple gauge group $G$ and the tensor fields transform in a non-trivial representation of $G$. The bosonic sector of this construction is based on the tensor hierarchy formalism \cite{deWit:2005hv} and it implies that, up to a number of tensor multiplets which are singlets of $G$, the system describes an off-shell Yang-Mills multiplet coupled to an off-shell massive tensor multiplet. The case of a single massive tensor, singlet under $G$, has already been constructed by other methods in \cite{D'Auria:2004sy,Louis:2004xi}, and partial results have been obtained for the nonabelian case in \cite{Nishino:2012sm}. In addition to an off-shell invariant containing the kinetic terms, we construct two additional off-shell invariants, one of which contains a mass term and another containing a topological ``theta'' term for the 2-form potential. We find that the elimination of the Yang-Mills multiplet auxiliary fields leads to an infinite number of terms in even powers of the tensor multiplet scalar fields. This nonpolynomial structure is due to the nonabelian nature of the massive tensors, and these terms disappear in the abelian case. We also show how the full off-shell $N=1, D=4$ supersymmetry transformations for the tensor-Yang-Mills system follow from a similar system in $D=6$ with $N=(1,0)$ supersymmetry \cite{Samtleben:2011fj,Samtleben:2012mi} upon dimensional reduction on a 2-torus followed by a truncation.

It is known that a massive tensor field in $4D$ is on-shell dual to a massive vector field. This dualization has been carried out for the system where all 2-form potentials are singlets of $G$ \cite{D'Auria:2004sy,Louis:2004xi}. Here, we generalize the dualization procedure to the case in which the 2-form potentials are in a nontrivial representation of $G$, thereby ending up with the couplings of on-shell nonabelian massive vector multiplets to off-shell Yang-Mills multiplets, which we will refer to as the supersymmetric Proca-Yang-Mills system.

It is interesting to compare this model with the supersymmetric Yang-Mills-Higgs system in which the massive vector multiplet arises by means of Higgs mechanism. The  latter models contain also massive scalar multiplets. We show that these can't be truncated  while preserving supersymmetry. Thus, the Proca-Yang-Mills system presented here provides an alternative Higgsless model for a description of massive vectors coupled to Yang-Mills. However, the tree-level scattering amplitude grows with energy for the longitudinal components of the massive vectors which means break-down in unitarity at high energies, unlike the situation in the Yang-Mills-Higgs model where the additional couplings of the Higgs scalars restore unitarity \cite{Bell:1973ex,LlewellynSmith:1973ey,Cornwall:1974km}. Motivated by an attempt to fix the tree-level unitarity problem in a bosonic Higgsless model where new couplings involving a 2-form potential are considered \cite{Lahiri:2011ic}, in Appendix B we extend the four parameter supersymmetric tensor-Yang-Mills system described in Section 3 by introducing two new off-shell invariants, which is possible if we take the 2-form potential in the adjoint representation. In this extended model new couplings relevant to the unitarity issue arise though a further study is required to determine whether they are sufficient to restore tree-level unitarity.


\section{The tensor hierarchy and supersymmetry}


Consider the coupling of $n_V$ vector and $n_T$ tensor off-shell $N=1, D=4$ supermultiplets
consisting of fields
\bea
&& \{ A_\mu^r, \lambda^r,D^r \}\ , \qquad r=1,...,n_V\ ,
\nonumber\\
&& \{ B_{\mu\nu}^I, \chi^I,\phi^I \}\ ,\qquad I=1,...,n_T\ ,
\eea
where the spinors are Majorana. The full covariant non-abelian field strengths are \cite{deWit:2005hv}
\begin{eqnarray}
{\cal F}_{\mu\nu}^r &\equiv&
2 \partial_{[\mu} A_{\nu]}^r - f_{st}{}^r A_\mu^s A_\nu^t + h^r{}_I\,B_{\mu\nu}^I
\;,\nonumber\\[.5ex]
{\cal H}_{\mu\nu\rho}^I &\equiv& 3 D_{[\mu} B_{\nu\rho]}^I +
6 \, d^I{}_{rs}\,  A_{[\mu}^r \partial^{\vphantom{r}}_\nu A_{\rho]}^s
- 2 f_{pq}{}^s d^I{}_{rs}\, A_{[\mu}^r A_\nu^p A_{\rho]}^q\ ,
\label{defF}
\end{eqnarray}
in terms of the antisymmetric structure constants $f_{st}{}^r=f_{[st]}{}^r$,
a symmetric $d$-symbol $d^I{}_{rs}=d^I{}_{(rs)}$, and the tensors $h^r{}_I$
inducing general St\"uckelberg-type couplings among forms of different degree.
We use canonical dimensions such that a $p$-form has mass dimension $p$
and as a result all constant tensors $f_{st}{}^r$, $d^I{}_{rs}$ and $h^r{}_I$, are dimensionless.
The covariant derivatives are defined as $D_\mu \equiv \partial_\mu - A_\mu^r X_r$ with
an action of the gauge generators $X_r$ on the different fields
given by $X_r \cdot \Lambda^s \equiv - (X_{r})_{t}{}^s \Lambda^t$,
$X_r \cdot \Lambda^I \equiv - (X_{r})_{J}{}^I \Lambda^J$, etc.
The field strengths (\ref{defF}) are defined such that they transform covariantly under the
set of non-abelian gauge transformations \cite{deWit:2005hv}
\begin{eqnarray}
\delta A_\mu^r &=& D_\mu \Lambda^r - h^r{}_I \Lambda_\mu^I
\;,\nonumber\\[.5ex]
\delta B_{\mu\nu}^I &=& 2 d^I{}_{rs}\,A_{[\mu}^r \,\delta A_{\nu]}^s2 + D_{[\mu} \Lambda_{\nu]}^I -2\, d^I{}_{rs} \,\Lambda^r {\cal F}_{\mu\nu}^s\ .
\label{gaugesym}
\end{eqnarray}
This vector/tensor gauge system is completely defined by the choice of the
invariant tensors $h^r{}_I$, $d^I{}_{rs}$, and $f_{rs}{}^t$.
Consistency of the tensor hierarchy, i.e.\ covariance of the field strengths (\ref{defF})
requires that the gauge group generators in the various representations are given by \cite{deWit:2005hv}
\begin{eqnarray}
(X_{r})_{s}{}^t &=&(X^{\rm V}_{r})_{s}{}^t~\equiv~ -f_{rs}{}^t + h^t{}_I\,d^I{}_{rs}\ ,
\nonumber\\
(X_{r})_{I}{}^J &=& (X^{\rm T}_{r})_{I}{}^J~\equiv~ 2\, d^J{}_{rs} h^s{}_I\ .
\label{genpar}
\end{eqnarray}
Further constraints follow from closure of the algebra (or generalized Jacobi identities)
$[X_r, X_s] = -(X_{r})_{s}{}^t \,X_t$ and gauge invariance of the tensors $h^r{}_I$ and $d^I{}_{rs}$.
Upon the use of these invariance conditions together with (\ref{genpar}), the generalized Jacobi identities take the equivalent form \cite{deWit:2005hv,Samtleben:2011fj,Samtleben:2012mi}
\begin{eqnarray}
{}
f_{rs}{}^t h^r{}_I - d^J{}_{rs}\,h^t{}_J h^r{}_I &=& 0\ ,
\label{c2}
\w2
f_{[pq}{}^u f_{r]u}{}^s - \frac13 h^s{}_I\, d^{I}{}_{u[p} f_{qr]}{}^u &=& 0\ ,
\label{c3}
\w2
h^s{}_I h^r{}_K \, d^J{}_{rs} &=& 0 .
\label{c4}
\end{eqnarray}

For an arbitrary matrix $h^r{}_I$, we can
choose a basis in the space of vector and two-form tensor fields according to
a split $A_\mu^r \longrightarrow \{A_\mu^\alpha , A_\mu^a\}$ and
$B_{\mu\nu}{}^I \longrightarrow \{B_{\mu\nu}{}_{a'} , B_{\mu\nu}{}^a\}$,
such that the matrix $h^r{}_I$ takes diagonal form \cite{Samtleben:2012mi}
\begin{eqnarray}
h^r{}_I &=&
 \left(
\begin{array}{cc}
h^\alpha{}^{b'} & h^\alpha{}_b \\
h^{a}{}^{b'} & h^a{}_b
\end{array}
\right)
~=~
\left(
\begin{array}{cc}
0 & 0 \\
0 & \delta^a_b
\end{array}
\right)\ ,
\label{r1}
\end{eqnarray}
with indices $a=1, \dots, {\rm rank}(h)$\, and indices $\alpha$, and $a'$
labeling the complement of the spaces of vector and tensor fields, respectively.
In this basis the constraints (\ref{c2}) -- (\ref{c3}) translate as follows:
the components $f_{\alpha\beta}{}^\gamma$ are the structure constants of a
Lie algebra $\mathfrak{g}$, satisfying standard Jacobi identities. Moreover,
we find that \cite{Samtleben:2012mi}
\begin{eqnarray}
 f_{ab}{}^c &=& 0\ ,\quad f_{ab}{}^\alpha=0\ ,\quad
f_{a\alpha}{}^\beta = 0\ ,\qquad f_{\alpha a}{}^b =-\frac12\,(T_\alpha)_a{}^b\ ,
\nn\\
d^c{}_{ab} &=& 0\ ,\qquad d^b{}_{\alpha a} = \frac12\,(T_\alpha)_a{}^b\ ,\quad
\label{r4}
\end{eqnarray}
where $T_\alpha$ are the generators of the Lie algebra $\mathfrak{g}$
in some representation ${\cal R}$, i.e.\
$[T_\alpha, T_\beta] = f_{\alpha\beta}{}^\gamma \,T_\gamma$,
with dimension ${\rm dim} \,{\cal R}={\rm rank}(h)$\ .
Together, we deduce that the generators (\ref{genpar}) take the form \cite{Samtleben:2012mi}
\begin{eqnarray}
X_\alpha^{\rm V} &=&
 \left(
\begin{array}{cc}
-f_{\alpha\beta}{}^\gamma &\;\;\; -f_{\alpha\beta}{}^b +d^b{}_{\alpha\beta} \\
0 & (T_\alpha)_a{}^b
\end{array}
\right)\ ,
\qquad X_a^{\rm V} ~=~ 0\ ,
\label{genV}
\end{eqnarray}
in the vector sector, and
\begin{eqnarray}
X_\alpha^{\rm T} &=&
 \left(
\begin{array}{cc}
 0 & 0  \\
2d_{b'}{}_{\alpha a}&(T_\alpha)_a{}^b
\end{array}
\right)\ , \qquad X_a^{\rm T} ~=~ 0\ ,
\label{genT}
\end{eqnarray}
in the tensor sector. Thus, the generalized Jacobi identity reduces to
$[X_\alpha, X_\beta] = f_{\alpha\beta}{}^\gamma \,X_\gamma$.
To summarize, we have solved the original system of constraints by an explicit choice of basis and without any loss of generality with \eq{r1} and \eq{r4}, where all non-vanishing tensors are invariant under the action of the generators (\ref{genV}) and (\ref{genT}), forming the Lie algebra $\mathfrak{g}$  with structure constants $f_{\alpha\beta}{}^\gamma$.

So far the solution of the constraints reviewed above is general. At this point, we will restrict to the case of a {\em semi-simple} Lie algebra ${\mathfrak{g}}$ and {\em non-trivial} representations $T_\alpha$.
In this case, by proper choice of basis, the matrices $X_\alpha$ from (\ref{genV}), (\ref{genT}) can be taken to be block-diagonal, i.e.\
\begin{eqnarray}
f_{\alpha\beta}{}^b &=&0 ~=~d^b{}_{\alpha\beta}\ ,\qquad
d_{b'}{}_{\alpha a}~=~0\ ,
\label{css}
\end{eqnarray}
and the generators take the form
\begin{eqnarray}
X_\alpha^{\rm V} &=&
 \left(
\begin{array}{cc}
-f_{\alpha\beta}{}^\gamma & 0 \\
0 & (T_\alpha)_a{}^b
\end{array}
\right)\ ,
\qquad
X_\alpha^{\rm T} ~=~
 \left(
\begin{array}{cc}
 0& 0 \\
0 & (T_\alpha)_a{}^b
\end{array}
\right)
\end{eqnarray}
which shows that the representation ${\cal R}'$ carried by the the 2-form potentials $B_{\mu\nu a'}$ is trivial.

In summary, the only non-vanishing components of the $h,f,d$ tensors are
\begin{eqnarray}
 f_{\alpha\beta}{}^\gamma\ ,\quad h^a{}_b=\delta^a_b\ , \quad
 f_{\alpha a}{}^b =-\frac12\,(T_\alpha)_a{}^b\ ,\quad
 d^b{}_{\alpha a} = \frac12\,(T_\alpha)_a{}^b\ ,\quad
d_{c'\alpha\beta}=d_{c'}\eta_{\alpha\beta}\ ,
\label{a1}
\end{eqnarray}
where $f_{\alpha\beta}{}^\gamma$ and $\eta_{\alpha\beta}$ are the structure constants
and Cartan-Killing form of a semi-simple Lie algebra $\mathfrak{g}$, respectively,
and $d_{c'}$ are arbitrary constants. The resulting model has the fields
\begin{eqnarray}
(A_\mu^\alpha, \lambda^\alpha,D^\alpha)\ ,\qquad
(A_\mu^a, \lambda^{a}, D^a)\ ,\qquad  (B_{\mu\nu}^a, \chi^{a}, \phi^a)\ ,
\label{fc1}
\end{eqnarray}
and a set of gauge singlet tensor multiplet fields $(B_{\mu\nu a'}, \chi_{a'}, \phi_{a'})$ where the indices $(\alpha, a,a')$ label  the adjoint representation, an irreducible representation ${\cal R}$ and singlets of $\mathfrak{g}$, respectively. The field equations for the singlet tensor multiplet of fields can be easily deduced from the results for non-abelian tensor multiplets, and therefore we shall leave them out.

The explicit bosonic field strengths are now given by\footnote{
In the particular case, when ${\cal R}$ is the adjoint representation, this gauge structure was considered in~\cite{Chu:2011fd}.}
\begin{eqnarray}
{\cal F}_{\mu\nu}^\alpha &=&
2 \partial_{[\mu} A_{\nu]}^\alpha - f_{\beta\gamma}{}^\alpha A_\mu^\beta A_\nu^\gamma ~\equiv~ { F}_{\mu\nu}^\alpha\ ,
\label{def1}\\[.5ex]
{\cal F}_{\mu\nu}^a &=&  2 D_{[\mu} A_{\nu]}^a -  (T_{\alpha})_{b}{}^a A_{[\mu}^\alpha A_{\nu]}^b
+ B_{\mu\nu}^a ~\equiv~ {\mathcal B}_{\mu\nu}^a\ ,
\label{def2}\\[.5ex]
{\cal H}_{\mu\nu\rho}^a &=& 3 D_{[\mu} {\mathcal B}_{\nu\rho]}^a\ .
\label{FBH}
\end{eqnarray}

The vector and tensor gauge-transformations are given by
\begin{eqnarray}
\delta A^\alpha_\mu&=&D_\mu\Lambda^\alpha\ ,
\nonumber\\
\delta A^a_\mu&=&D_\mu\Lambda^a-\Lambda^a_\mu\ ,
\nonumber\\
\delta B^a_{\mu\nu}&=&2D_{[\mu}\Lambda^a_{\nu]}-T_{\alpha b}{}^a\left(\Lambda^\alpha\mathcal B^b_{\mu\nu}+\Lambda^bF^\alpha_{\mu\nu}-A^\alpha_{[\mu}\delta A^b_{\nu]}
-A^b_{[\mu}\delta A^\alpha_{\nu]}\right)\ .
\end{eqnarray}
This implies that the massive tensor field $\mathcal B$ transforms as a Yang-Mills covariant field strength
\begin{equation}
\delta\mathcal B^a_{\mu\nu}=-T_{\alpha b}{}^a\Lambda^\alpha\mathcal B^b_{\mu\nu}\ .
\end{equation}
The scalar fields $\phi^a$ transform the same way as $\mathcal B^a_{\mu\nu}$ under the gauge transformations, and their covariant derivative takes the form
\be
D_\mu\phi^a=\partial_\mu\phi^a+A_\mu^\alpha T_{\alpha b}{}^a\phi^b\ .
\label{eq:Dphi}
\ee

Turning to supersymmetry, the results summarized above for the bosonic tensor-Yang-Mills system shows that the coupling problem we started with amounts to constructing the coupling of the following multiplets
\bea
&&\mbox{Off-shell Yang-Mills Multiplet:}\qquad\qquad\ \ \ (A_\mu^\alpha,\lambda^\alpha,D^\alpha)\ ,
\w2
&& \mbox{Off-shell\ Massive\ Tensor\ Multiplet:}
\qquad\quad  ({\mathcal B}_{\mu\nu}^a, \chi^{a}, \phi^a, \lambda^{a}, D^a)\ .
\eea
Note that the tensor field $B_{\mu\nu}^a$ has eaten the vector $A_\mu^a$ to become the massive tensor $\mathcal B_{\mu\nu}^a$ through the St\"uckelberg mechanism.

The supersymmetry transformation rules of the off-shell Yang-Mills multiplet are well known and are given by
\begin{eqnarray}
\delta A_\mu^\alpha&=&-{\bar\epsilon}\gamma_\mu\lambda^\alpha\ ,
\nonumber\\
\delta\lambda^\alpha&=&\frac{1}{8}F_{\mu\nu}^\alpha\,\gamma^{\mu\nu}\epsilon+\frac{i}{2}D^\alpha\,\gamma_5\epsilon
\nonumber\\
\delta D^\alpha&=&\frac{i}{2}{\bar\epsilon}\gamma_5 {\slashed D} \lambda^\alpha\ .
\label{YM}
\end{eqnarray}
As for the supersymmetry transformation rules for the massive tensor multiplet, so far they have been determined fully only for an abelian version of the multiplet. Here we present the full result for the nonabelian case, by which we mean that all the members of the multiplet, including the tensor field, carry the same and nontrivial representation of the semi-simple gauge group $G$. In addition to employing the Yang-Mills covariant field strengths described above, and covariantizing derivatives of the fermions as well by employing the Yang-Mills gauge field $A_\mu^\alpha$, we find that we need higher order fermionic terms in the transformation rule for the tensorino in which the Yang-Mills gaugino fields appear explicitly. To determine the full result, we establish the off-shell closure of the supersymmetry algebra on all the members of the massive tensor-Yang-Mills multiplet, and we thus find\footnote{In our conventions $\eta_{\mu\nu}={\rm diag}(-1,+1,+1,+1)$ and
$\gamma_{\mu\nu\rho\sigma}=i\varepsilon_{\mu\nu\rho\sigma}\gamma_5$.  A crucial Fierz identity is
$(C\gamma_\mu)_{\alpha(\beta}\,(C\gamma^\mu)_{\gamma\delta)}=0$, where $C$ is the anti-symmetric charge conjugation matrix with $C^2=-1$.
More generally we have
$
\psi{\bar\chi}
=
-\frac{1}{4} {\bar\psi}\chi
-\frac{1}{4}{\bar\psi}\gamma_5\chi\,\gamma_5
+\frac{1}{4}{\bar\psi}\gamma^\mu\chi\,\gamma_\mu
-\frac{1}{8}{\bar\psi}\gamma^{\mu\nu}\chi\,\gamma_{\mu\nu}
+\frac{1}{4}{\bar\psi}\gamma^\mu\gamma_5\chi\,\gamma_\mu\gamma_5\,.$
}
\begin{eqnarray}
\delta\mathcal B^a_{\mu\nu}&=&2{\bar\epsilon}\gamma_{[\mu}D_{\nu]}\lambda^a-\bar\epsilon\gamma_{\mu\nu}\chi^a
\nonumber\\
\delta\phi^a&=&{\bar\epsilon}\chi^a
\nonumber\\
\delta\chi^a&=&\frac{1}{24}\mathcal H^a_{\mu\nu\rho}\,\gamma^{\mu\nu\rho}\epsilon
+\frac{1}{4}{\slashed D} \phi^a\,\epsilon
-\frac{1}{2}T_{\alpha b}{}^a\,(\gamma^\mu\lambda^{(\alpha})\,{\bar\epsilon}\gamma_\mu\lambda^{b)}
+\frac{1}{4}T_{\alpha b}{}^a\,(\gamma^\mu\gamma_5\epsilon)\,{\bar\lambda}^\alpha\gamma_\mu\gamma_5\lambda^b
\nonumber\\
\delta\lambda^a&=&\frac{1}{8}\mathcal B^a_{\mu\nu}\,\gamma^{\mu\nu}\epsilon
+\frac{i}{2}D^a\,\gamma_5\epsilon+\frac{1}{4}\phi^a\,\epsilon
\nonumber\\
\delta D^a&=&\frac{i}{2}{\bar\epsilon}\gamma_5 {\slashed D}\lambda^a
-\frac{i}{2}{\bar\epsilon}\gamma_5\chi^a\ .
\label{eq:tensorYM}
\end{eqnarray}
Though we will only work in terms of $\mathcal B_{\mu\nu}^a$ we give for completeness the supersymmetry transformation of $B^a_{\mu\nu}$ and $A^a_\mu$:
\begin{eqnarray}
\delta B^a_{\mu\nu}&=&
-{\bar\epsilon}\gamma_{\mu\nu}\chi^a
-T_{\alpha b}{}^aA^\alpha_{[\mu}\,{\bar\epsilon}\gamma_{\nu]}\lambda^b
-T_{\alpha b}{}^aA^b_{[\mu}\,{\bar\epsilon}\gamma_{\nu]}\lambda^\alpha\nonumber\\
\delta A^a_\mu&=&-{\bar\epsilon}\gamma_\mu\lambda^a\ .
\end{eqnarray}

In Appendix A, we describe an alternative derivation of (\ref{eq:tensorYM}) by a dimensional reduction from six dimensions, followed by a consistent truncation, of $N=(1,0)$ superconformal tensor-Yang-Mills system described in \cite{Samtleben:2011fj,Samtleben:2012mi}. It is also useful to note that in the abelian limit of the supersymmetry transformation \eq{eq:tensorYM}, the $\epsilon\lambda^2$ terms in $\delta\chi^a$ are absent and all the covariant derivatives become ordinary partial derivatives.


\section{Action for nonabelian massive tensor-Yang-Mills system}


To write an action we must introduce also a massive tensor multiplet transforming in the conjugate representation, unless the representation happens to be self-conjugate. These fields will be denoted by lower indices, and thus we have
\be
\mbox{Off-shell\ Massive\ Conjugate\ Tensor\ Multiplet:}
\qquad\quad  ({\mathcal B}_{\mu\nu a}, \chi_a, \phi_a, \lambda_a, D_a)\ .
\label{cmtv}
\ee
The supersymmetry transformations of these are obtained from those given above by hermitian conjugation according to the rules:
\be
V^a = (V_a)^\dagger\ ,\qquad (T_{\alpha b}{}^a)^\dagger= -T_{\alpha a}{}^b\ ,
\ee
where $V_a$ is a generic field. We now wish to construct a Lagrangian for the massive tensor coupled to Yang-Mills. The off-shell $N=1$ super Yang-Mills Lagrangian is the standard one. We construct three new off-shell invariants involving the massive tensor multiplet and its interaction with the off-shell Yang-Mills. The final result is
\begin{equation}
\mathcal L =\mathcal L_{\mathrm{bos}}+\mathcal L_{\mathrm{ferm}}\ ,
\label{master}
\end{equation}
where
\begin{eqnarray}
\mathcal L_{\mathrm{bos}} &=&-\frac{1}{4}F^\alpha_{\mu\nu}F_\alpha^{\mu\nu}+2D^\alpha D_\alpha
-\frac{1}{6}\mathcal H^a_{\mu\nu\rho}\mathcal H_a^{\mu\nu\rho}
-D_\mu\phi^aD^\mu\phi_a
\nn\w2
&&{}
-\frac{1}{2}T_{\alpha b}{}^a \left(\phi^b\mathcal B^{\mu\nu}_a \mathcal F^\alpha_{\mu\nu}
+8\phi_aD^bD^\alpha -(a\leftrightarrow b) \right)
\nn\w2
&&{}
-\frac{1}{2} m^2 \left(\mathcal B^a_{\mu\nu}\mathcal B_a^{\mu\nu}+2\phi^a\phi_a-8D^aD_a\right)
\nn\w2
&&{}
-\frac{1}{8} \theta m^2\left(\varepsilon^{\mu\nu\rho\sigma}\mathcal B^a_{\mu\nu}\mathcal B_{\rho\sigma\,a}
-16\phi^aD_a + h.c.\right)\ ,
\label{eq:bos}
\end{eqnarray}
and
\begin{eqnarray}
\mathcal L_{\mathrm{ferm}} &=&-2{\bar\lambda}^\alpha{\slashed D}\lambda_\alpha
-4{\bar\chi}^a {\slashed D} \chi_a-4m^2 {\bar\lambda}^a{\slashed D}\lambda_a
\nn\w2
&&
+4m^2\left( {\bar\lambda}^a\chi_a +{\bar\lambda}_a\chi^a \right)
+4i\theta m^2\left( {\bar\lambda}^a\gamma_5\chi_a +{\bar\lambda}_a\gamma_5\chi^a\right)
\nn\w2
&&
-T_{\alpha b}{}^a
\Big[
F^\alpha_{\mu\nu}\,{\bar\lambda}^b\gamma^{\mu\nu}\chi_a
+\mathcal B^b_{\mu\nu}\,{\bar\lambda}^\alpha\gamma^{\mu\nu}\chi_a
+\frac{1}{3}\mathcal H_{a\,\mu\nu\rho}\,{\bar\lambda}^\alpha\gamma^{\mu\nu\rho}\lambda^b
\nonumber\\
&&{}
-6\phi^b\,{\bar\chi}_a\lambda^\alpha
+2D_\mu\phi_a\,{\bar\lambda}^\alpha\gamma^\mu\lambda^b
-4\phi_a\,{\bar\lambda}^b {\slashed D} \lambda^\alpha
+4iD^b\,{\bar\lambda}^\alpha\gamma_5\chi_a
\nonumber\\
&&{}
+4iD^\alpha\,{\bar\lambda}^b\gamma_5\chi_a
-(a\leftrightarrow b)
\Big]
\nonumber\\
&&{}
-2T_{(\alpha b}{}^aT_{\beta) a}{}^c
\left(
2{\bar\lambda}^b\gamma_5\gamma^\mu\lambda^\beta\,{\bar\lambda}_c\gamma_\mu\gamma_5\lambda^\alpha
+{\bar\lambda}^b\gamma_5\gamma^\mu\lambda_c\,{\bar\lambda}^\beta\gamma_\mu\gamma_5\lambda^\alpha
\right)\ ,
\label{eq:ferm}
\end{eqnarray}
where $m\neq0$ is a mass parameter and $\theta$ is a dimensionless constant. The parts of the total Lagrangian consisting of the terms that depend only on $(A_\mu^\alpha,\lambda^\alpha,D^\alpha)$, those proportional to $m^2$ and $\theta m^2$ and the remainder, are separately off-shell supersymmetric and Yang-Mills gauge invariant. It should not be too difficult to write the action in superspace but we will not attempt to do this here. The abelian version of this action (coupled to additional vectors and scalars) was considered in \cite{Louis:2004xi}.

It is also worth observing that if we set $m^2=0$, then the total Lagrangian does not contain a kinetic term for $\lambda^a$ and the ${\mathcal B}$ field equation
\begin{equation}
D^\rho\mathcal H^a_{\mu\nu\rho}-T_{\alpha b}{}^a\phi^bF^\alpha_{\mu\nu}=0
\end{equation}
gives an unwanted constraint, as can be seen by taking its divergence, which yields
\footnote{Note that using the equation of motion for the Yang-Mills gauge field,
\begin{eqnarray}
D_\nu F_\alpha^{\mu\nu}
+T_{\alpha b}{}^a
\Big(
D_\nu(\phi^b\mathcal B^{\mu\nu}_a-\phi_a\mathcal B^{\mu\nu\,b})
+\phi^bD^\mu\phi_a
-\phi_aD^\mu\phi^b
+\frac{1}{2}\mathcal B^b_{\nu\rho}\mathcal H_a^{\mu\nu\rho}
-\frac{1}{2}\mathcal B_{\nu\rho\,a}\mathcal H^{\mu\nu\rho\,b}
\Big)
=0\ ,
\end{eqnarray}
does not help.}
\begin{equation}
T_{\alpha b}{}^a
\left(
F^{\nu\rho\,\alpha}\mathcal H^b_{\mu\nu\rho}
-2D^\nu(\phi^bF^\alpha_{\mu\nu})
\right)
=0\ .
\end{equation}
This problem is avoided by letting $m^2\ne 0$, in which case the ${\mathcal B}$ field equation now implies a constraint on $D^\mu\mathcal B_{\mu\nu}$ which is a typical subsidiary condition for massive fields (recall the standard Proca equation which implies that the vector is divergence free). Taking $m^2\ne 0$ also furnishes a kinetic term for $\lambda^a$.


\section{Integrating out the auxiliary fields}


In this section we will show that when integrating out the auxiliary fields in \eq{master} we obtain an on-shell supersymmetric action which involves an infinite expansion in even powers of $\phi$. The equations of motion of the $D$'s are
\begin{eqnarray}
m^2D^a&=&-\frac12\theta m^2\phi^a-T_{\alpha b}{}^a\left(\phi^b\,D^\alpha+i{\bar\lambda}^\alpha\gamma_5\chi^b\right)\ ,
\label{aux1}\\
m^2D_a&=&-\frac12\theta m^2\phi_a+T_{\alpha a}{}^b\left(\phi_b\,D^\alpha+i{\bar\lambda}^\alpha\gamma_5\chi_b\right)\ ,
\label{aux2}\\
D_\alpha&=&T_{\alpha b}{}^a\left(\phi_a\,D^b-\phi^b\,D_a+i{\bar\lambda}^b\gamma_5\chi_a
-i{\bar\lambda}_a\gamma_5\chi^b\right)\ .
\label{aux3}
\end{eqnarray}
Using the first two equations in the last one gives
\bea
&&
D^\alpha=
-\frac{i}{m^2}\,{\mathcal G}^{\alpha\beta}T_{\beta b}{}^a
\left(
\phi_a\,T_{\gamma c}{}^b{\bar\lambda}^\gamma\gamma_5\chi^c
+\phi^b\,T_{\gamma a}{}^c{\bar\lambda}^\gamma\gamma_5\chi_c
-m^2{\bar\lambda}^b\gamma_5\chi_a
+m^2{\bar\lambda}_a\gamma_5\chi^b \right)\ ,\qquad
\label{aux4}
\eea
where
\boxedeq{{\mathcal G}^{\alpha\beta}= \left(\delta_{\alpha\beta}
+\frac{2}{m^2}\phi^b\,(T_{(\alpha}T_{\beta)})_b{}^a\phi_a\right)^{-1} \label{eq:aux2}}

\bigskip

The representation matrices are defined such that $T_{(\alpha}T_{\beta)}=\frac{1}{2 d_R}\delta_{\alpha\beta}+\frac{i}{2}d_{\alpha\beta\gamma}T^\gamma$, where $d_R$ is the dimension of the representation, so that for an $SO(N)$ gauge group for example, for which $d_{\alpha\beta\gamma}=0$, we have
${\mathcal G}_{\alpha\beta}=(1+ (m^2 d_R)^{-1} \phi^2)^{-1}\,\delta_{\alpha\beta}$.

Using the expressions for the auxiliary fields in \eq{aux1}-\eq{eq:aux2} in the Lagrangian (\ref{master}) we obtain, after some algebra, the on-shell Lagrangian
\begin{equation}
\mathcal L=\mathcal L_{\mathrm b}+\mathcal L_{2\mathrm f}+\mathcal L_{4\mathrm f}\ ,
\label{MTV}
\end{equation}
where the bosonic terms are
\begin{eqnarray}
\mathcal L_{\mathrm b}&=&
-\frac{1}{6}\mathcal H^a_{\mu\nu\rho}\mathcal H_a^{\mu\nu\rho}
-\frac{1}{4}F^\alpha_{\mu\nu}F_\alpha^{\mu\nu}
-D_\mu\phi^aD^\mu\phi_a
-\frac12 m^2\mathcal B^a_{\mu\nu}\mathcal B_a^{\mu\nu}
-M^2\phi^a\phi_a
\nonumber\\
&&{}
-\frac14\theta m^2\varepsilon^{\mu\nu\rho\sigma}\mathcal B^a_{\mu\nu}\mathcal B_{\rho\sigma\,a}
-\frac{1}{2}T_{\alpha b}{}^a\left(\phi^bF^\alpha_{\mu\nu}\mathcal B^{\mu\nu}_a-\phi_aF^\alpha_{\mu\nu}\mathcal B^{\mu\nu\,b}\right)\ ,
\end{eqnarray}
with the $M^2$ defined by 
\boxedeq{M^2=m^2(1+\theta^2)\,.}
The quadratic fermion terms are
\begin{eqnarray}
\mathcal L_{2\mathrm f}&=&
-4{\bar\chi}^a {\slashed D} \chi_a
-4m^2{\bar\lambda}_a {\slashed D} \lambda^a
-2{\bar\lambda}^\alpha {\slashed D}\lambda_\alpha
+4m^2\left[{\bar\lambda}^a(1+i\theta\gamma_5)\chi_a+{\bar\lambda}_a(1+i\theta\gamma_5)\chi^a\right]
\nonumber\\
&&{}
-T_{\alpha b}{}^a
\Big[
-6\phi^b\,{\bar\chi}_a\lambda^\alpha
+\frac{1}{3}\mathcal H_{a\,\mu\nu\rho}\,{\bar\lambda}^\alpha\gamma^{\mu\nu\rho}\lambda^b
+\mathcal B^b_{\mu\nu}\,{\bar\lambda}^\alpha\gamma^{\mu\nu}\chi_a
+F^\alpha_{\mu\nu}\,{\bar\lambda}^b\gamma^{\mu\nu}\chi_a
\nonumber\\
&&{}
+2D_\mu\phi_a\,{\bar\lambda}^\alpha\gamma^\mu\lambda^b
-4\phi_a\,{\bar\lambda}^b {\slashed D}\lambda^\alpha
+2i\theta\phi_a\,{\bar\lambda}^\alpha\gamma_5\chi^b
-(a\leftrightarrow b)
\Big]
\end{eqnarray}
and the quartic fermion terms are
\begin{eqnarray}
\mathcal L_{4\mathrm f}&=&
-2(T_{(\alpha}T_{\beta)})_b{}^a
\left(
2{\bar\lambda}^b\gamma_5\gamma^\mu\lambda^\beta\,{\bar\lambda}_a\gamma_\mu\gamma_5\lambda^\alpha
+{\bar\lambda}^b\gamma_5\gamma^\mu\lambda_a\,{\bar\lambda}^\beta\gamma_\mu\gamma_5\lambda^\alpha
\right)
-\frac{4}{m^2}(T_\alpha T_\beta)_b{}^a\,{\bar\lambda}^\alpha\gamma_5\chi^b\,{\bar\lambda}^\beta\gamma_5\chi_a
\nonumber\\
&&{}
+\frac{2}{m^4}{\mathcal G}^{\beta\delta}\,T_{\beta b}{}^a T_{\delta c}{}^d
\Big(
T_{\alpha a}{}^e\phi^b\,{\bar\lambda}^\alpha\gamma_5\chi_e
+T_{\alpha e}{}^b\phi_a\,{\bar\lambda}^\alpha\gamma_5\chi^e
+m^2{\bar\lambda}_a\gamma_5\chi^b
-m^2{\bar\lambda}^b\gamma_5\chi_a
\Big)
\nonumber\\
&&
\cdot\Big(
T_{\gamma f}{}^c\,\phi_d\,{\bar\lambda}^\gamma\gamma_5\chi^f
+T_{\gamma d}{}^f\,\phi^c\,{\bar\lambda}^\gamma\gamma_5\chi_f
+m^2{\bar\lambda}_d\gamma_5\chi^c
-m^2{\bar\lambda}^c\gamma_5\chi_d
\Big)\ ,
\end{eqnarray}
where ${\mathcal G}^{\alpha\beta}$ defined in \eq{eq:aux2} is the source of nonpolynomial interactions of the scalar fields with the fermions. The elimination of the auxiliary fields have given rise to the $\theta^2$ dependent term in ${\mathcal L}_{\mathrm b}$ and all the terms proportional to inverse powers of $m$ in $\mathcal L_{2\mathrm f}$ and $\mathcal L_{4\mathrm f}$. All the terms of the form ${\mathcal G} \times \psi^2$ have canceled, where $\psi$ denotes the fermions.

The action of the Lagrangian \eq{MTV} is invariant under supersymmetry transformations \eq{YM} and \eq{eq:tensorYM}, with the substitutions \eq{aux1}-\eq{eq:aux2}. If we set  $\theta=0$ and truncate to cubic order in fields, the Lagrangian up to field redefinitions\footnote{The required field redefinitions are
$
\rho^i\rightarrow2\sqrt2\,i(m\,\lambda^a+m^{-1}\,T_{\alpha b}{}^a\phi^b\lambda^\alpha),\
\varphi\rightarrow\sqrt2\,\phi\,,\quad H_{\mu\nu}\rightarrow\sqrt2\,m\mathcal B_{\mu\nu}\,,\quad\chi\rightarrow2\sqrt2\,i\chi\,,\quad\lambda^I\rightarrow-2i\lambda^\alpha
$
together with $g\rightarrow m$, $D_\mu\rightarrow\frac{1}{m}D_\mu$.}
agrees with the Lagrangian given in \cite{Nishino:2012sm}, where the Lagrangian as well as the supersymmetry transformations were constructed only to this order.


\section{Dualization to a massive vector}


It is known that a single massive tensor is dual to a massive vector field. The dualization was described in superspace in \cite{Louis:2004xi}. Here we will perform the corresponding dualization in the non-abelian case.
The first step is to implement the fact that $\mathcal H=D\mathcal B$ by adding to the Lagrangian
\begin{equation}
\mathcal L_{\mathrm{constr.}}=
\frac16 M\varepsilon^{\mu\nu\rho\sigma}C^a_\mu(3D_\nu\mathcal B_{\rho\sigma\,a}-\mathcal H_{\nu\rho\sigma\,a})
+ h.c.\ ,
\end{equation}
where $C^a_\mu$ and $C_{\mu\,a}$ are Lagrange multiplier fields enforcing the constraint and the coefficient is chosen for later convenience. In the new Lagrangian
\begin{equation}
{\mathcal L'}={\mathcal L}+\mathcal L_{\mathrm{constr.}}\ ,
\label{L2}
\end{equation}
where ${\mathcal L}$ is given in \eq{master}, $\mathcal H$ is treated as an independent field. The equation of motion of $\mathcal B^a$ now reads
\boxedeq{
\mathcal B^a_{\mu\nu}
=
M^{-1}\left(\theta G^a_{\mu\nu}+ *G^a_{\mu\nu}\right)
- M^{-2}\,T_{\alpha b}{}^a\left(
\phi^bF^\alpha_{\mu\nu}
-\theta\phi^b*F^\alpha_{\mu\nu}
-2{\bar\lambda}^\alpha\gamma_{\mu\nu}(1-i\theta\gamma_5)\chi^b
\right)\ ,\label{dr}}
where $G=DC$ is the Yang-Mills covariant field strength of $C$ and the Hodge dual is given by $*G_{\mu\nu}=\frac{1}{2}\varepsilon_{\mu\nu\rho\sigma}G^{\rho\sigma}$. The equation of motion of $\mathcal H$ reads
\boxedeq{
\mathcal H^a_{\mu\nu\rho}=M\varepsilon_{\mu\nu\rho\sigma}C^{\sigma\,a}-2T_{\alpha b}{}^a\,{\bar \lambda}^\alpha\gamma_{\mu\nu\rho}\lambda^b\,.    \label{eq:HC} }
The off-shell Lagrangian \eq{master} as well as the on-shell one \eq{MTV} give the same equations for ${\mathcal B}_{\mu\nu}$ and $\mathcal H_{\mu\nu\rho}$. Therefore, we can substitute back these results in either Lagrangian. For the convenience of having the Yang-Mills system off-shell, we use the expressions for $\mathcal B$ and $\mathcal H$ in \eq{master}, together with the expression for $D^a$ \eq{aux1}, thereby obtaining the full Lagrangian for the on-shell super Proca system coupled to off-shell Yang-Mills given by:
\be
{\mathcal L'} ={\mathcal L'}_{\mathrm b}+{\mathcal L'}_{2\mathrm f}+{\mathcal L'}_{4\mathrm f}\ ,
\label{L22}
\ee
where the bosonic terms are
\begin{eqnarray}
{\mathcal L'}_{\mathrm b}&=&-\frac{1}{4}F^\alpha_{\mu\nu}F_\alpha^{\mu\nu}
-\frac{1}{2}G^a_{\mu\nu}\left(G^{\mu\nu}_a-\theta*G^{\mu\nu}_a\right)-D_\mu\phi^aD^\mu\phi_a
-M^2C^a_\mu C_a^\mu-M^2\phi^a\phi_a+2D^\alpha D_\alpha
\nonumber\\
&&{}
-\frac{1}{2M}T_{\alpha b}{}^a\,F_{\mu\nu}^\alpha\left(
\theta \phi^b G_a^{\mu\nu} -\phi_a*G^{\mu\nu b}
-\theta \phi_a G^{\mu\nu b} +\phi^b*G^{\mu\nu}_a \right)
\nonumber\\
&&{}
-\frac{1}{2M^2}(T_\alpha T_\beta)_b{}^a \phi_a\phi^b F^\alpha_{\mu\nu}
\Big(F^{\mu\nu\,\beta} -\theta *F^{\mu\nu\,\beta}\Big)
+\frac{4}{m^2}(T_\alpha T_\beta)_b{}^a\,\phi_a\,\phi^b\,D^\alpha\,D^\beta\ ,
\label{Lbos}
\end{eqnarray}
the quadratic fermion terms are
\begin{eqnarray}
{\mathcal L'}_{2\mathrm f}&=&
-2{\bar\lambda}^\alpha{\slashed D}\lambda_\alpha-4{\bar\chi}^a{\slashed D}\chi_a
-4m^2{\bar\lambda}_a{\slashed D}\lambda^a
+4m^2\left({\bar\lambda}^a(1+i\theta\gamma_5)\chi_a+{\bar\lambda}_a(1+i\theta\gamma_5)\chi^a\right)
\nonumber\\
&&{}
+\frac{1}{M}T_{\alpha b}{}^a
\Big(
iG_a^{\mu\nu}\,{\bar\lambda}^\alpha\gamma_{\mu\nu}\gamma_5(1-i\theta\gamma_5)\chi^b
-6M\phi_a\,{\bar\lambda}^\alpha\chi^b
-2iM\theta\phi_a\,{\bar\lambda}^\alpha\gamma_5\chi^b
\nonumber\\
&&{}
-2iM^2C_a^\mu\,{\bar\lambda}^\alpha\gamma_\mu\gamma_5\lambda^b
-2MD_\mu\phi_a\,{\bar\lambda}^\alpha\gamma^\mu\lambda^b
+4M\phi_a\,{\bar\lambda}^b{\slashed D}\lambda^\alpha
\nonumber\\
&&{}
-MF^\alpha_{\mu\nu}\,{\bar\lambda}^b\gamma^{\mu\nu}\chi_a
-4iMD^\alpha\,{\bar\lambda}^b\gamma_5\chi_a
-(a\leftrightarrow b)
\Big)
\nonumber\\
&&{}
+\frac{1}{M^2}(T_\alpha T_\beta)_b{}^a
\Big(
\phi_aF^\beta_{\mu\nu}\,{\bar\lambda}^\alpha\gamma^{\mu\nu}(1-i\theta\gamma_5)\chi^b
+\phi^bF^\alpha_{\mu\nu}\,{\bar\lambda}^\beta\gamma^{\mu\nu}(1-i\theta\gamma_5)\chi_a
\Big)
\nonumber\\
&&{}
+\frac{4i}{m^2}(T_\alpha T_\beta)_b{}^a
\Big(
\phi_a\,D^\beta\,{\bar\lambda}^\alpha\gamma_5\chi^b
+\phi^b\,D^\alpha\,{\bar\lambda}^\beta\gamma_5\chi_a
\Big)
\end{eqnarray}
and the quartic fermion terms are
\begin{eqnarray}
{\mathcal L'}_{4\mathrm f}&=&
-\frac{2}{M^2}(T_\alpha T_\beta)_b{}^a
\Big[
2(1+\theta^2){\bar\lambda}^\beta\gamma_5\chi_a\,{\bar\lambda}^\alpha\gamma_5\chi^b
+{\bar\lambda}^\alpha\gamma_{\mu\nu}(1-i\theta\gamma_5)\chi^b\,{\bar\lambda}^\beta\gamma^{\mu\nu}\chi_a
\\
&&{}
+M^2{\bar\lambda}^b\gamma_\mu\gamma_5\lambda^\alpha\,{\bar\lambda}_a\gamma^\mu\gamma_5\lambda^\beta
-M^2{\bar\lambda}^b\gamma^\mu\gamma_5\lambda^\beta\,{\bar\lambda}_a\gamma_\mu\gamma_5\lambda^\alpha
+M^2{\bar\lambda}^b\gamma_5\gamma^\mu\lambda_a\,{\bar\lambda}^\beta\gamma_\mu\gamma_5\lambda^\alpha
\Big]\ .\nonumber
\end{eqnarray}
Equation (\ref{eq:HC}) which expresses $\mathcal H$ in terms of $C$ can be used to determine the supersymmetry transformation of the massive vector field $C$. One finds
\begin{eqnarray}
\delta C^a_\mu
&=&
\frac{im^2}{M}{\bar\epsilon}\gamma_\mu\gamma_5(1+i\theta\gamma_5)\lambda^a
+\frac{i}{M}{\bar\epsilon}\gamma_5D_\mu\chi^a
+\frac{i}{M}T_{\alpha b}{}^a\phi^b\,{\bar\epsilon}\gamma_\mu\gamma_5\lambda^\alpha
\nonumber\\
&&{}
+\frac{i}{M}{\bar\epsilon}\gamma_\mu\gamma_5(\gamma^\nu D_\nu\chi^a-m^2\lambda^a+\ldots)\ ,
\label{dC}
\end{eqnarray}
where the last term is proportional to the equation of motion for $\chi$ and, since in the dualized model the supersymmetry algebra will be on--shell we will simply drop this term. The other supersymmetry transformations are obtained by using the expressions for $\mathcal B$ and $\mathcal H$ in terms of the vector field $C$ in the supersymmetry algebra (\ref{eq:tensorYM}). The supersymmetry transformations which leave this action invariant are therefore  the standard ones for the off-shell Yang-Mills fields (\ref{YM}) together with the on-shell supersymmetry transformation of the massive vector multiplet
\begin{eqnarray}
\delta C^a_\mu
&=&
\frac{im^2}{M}{\bar\epsilon}\gamma_\mu\gamma_5(1+i\theta\gamma_5)\lambda^a
+\frac{i}{M}{\bar\epsilon}\gamma_5D_\mu\chi^a
+\frac{i}{M}T_{\alpha b}{}^a\phi^b\,\bar\epsilon\gamma_\mu\gamma_5\lambda^\alpha
\nonumber\\
\delta\phi^a&=&{\bar\epsilon}\chi^a
\nonumber\\
\delta\chi^a&=&
\frac{i}{4}M C^a_\mu\,\gamma^\mu\gamma_5\epsilon
+\frac{1}{4}D_\mu\phi^a\,\gamma^\mu\epsilon
-\frac{1}{2}T_{\alpha b}{}^a\,(\gamma^\mu\lambda^{(\alpha})\,{\bar\epsilon}\gamma_\mu\lambda^{b)}
-\frac{1}{4}T_{\alpha b}{}^a\,(\gamma^\mu\gamma_5\epsilon)\,{\bar\lambda}^\alpha\gamma_\mu\gamma_5\lambda^b
\nonumber\\
\delta\lambda^a&=&
\frac{i}{8M}G^a_{\mu\nu}\,\gamma^{\mu\nu}\gamma_5(1-i\theta\gamma_5)\epsilon
-\frac{1}{8M^2}T_{\alpha b}{}^a\left(\phi^bF^\alpha_{\mu\nu}-2{\bar\lambda}^\alpha\gamma_{\mu\nu}\chi^b\right)\,\gamma^{\mu\nu}(1-i\theta\gamma_5)\epsilon
\nonumber\\
&&{}
-\frac{i}{2m^2}T_{\alpha b}{}^a\left(\phi^b\,D^\alpha+i{\bar\lambda}^\alpha\gamma_5\chi^b\right)\,\gamma_5\epsilon
+\frac{1}{4}\phi^a\,(1-i\theta\gamma_5)\epsilon
\ .
\end{eqnarray}
As a byproduct of the dualization of a massive tensor we have shown how to couple a massive vector, transforming in a non-trivial representation of the gauge-group $G$, to Yang-Mills. Note that this construction does not involve any scalar multiplets and is therefore quite different from the usual approach via the Higgs mechanism. A more detailed analysis of the differences will be presented in the next section. Note also that if we integrate out the auxiliary field of the Yang-Mills multiplet $D^\alpha$ using equation (\ref{aux4}) and (\ref{eq:aux2}) we get an infinite series of terms involving powers of the scalar $\phi$, both in the supersymmetry transformations and in the action. If we had started with a completely on-shell formulation we would therefore only have been able to construct the model order by order in these terms.

It should be possible to find a completely off-shell description of this system. To do this one should introduce the appropriate auxiliary fields to make the massive vector multiplet off-shell. These consist of one complex and one real scalar. It should then be possible to formulate this model in superspace. We leave this problem for the future.


\section{Comparison to supersymmetric Higgs model}


 The supersymmetric Proca-Yang-Mills in the previous section contains only the massive vector multiplet and the Yang-Mills multiplet, and in particular no scalar multiplets. Here we shall compare this model with the supersymmetric Yang-Mills-Higgs model which does contain scalar multiplets. We will see that, except in the abelian case, the truncation of the scalar multiplets without breaking supersymmetry is not possible.  
 Therefore the two models represent completely distinct ways of coupling massive vectors to super Yang-Mills. 
 
To be concrete we will analyze the specific example of breaking $SU(n+1)\rightarrow SU(n)\times U(1)$ ($n>1$) via a supersymmetric version of the Higgs mechanism, which arises for example in supersymmetric GUTs \cite{Dimopoulos:1981zb}. The starting point is the (renormalizable) Lagrangian
\begin{eqnarray}
\mathcal L_{\mathrm{Higgs}}&=&
\mathrm{Tr}\Big[
-\frac{1}{4}F_{\mu\nu}F^{\mu\nu}-2\bar\lambda{\slashed D}\lambda
+\left(\frac14 m\bar\psi\psi_L+\frac{1}{2}z\,\bar\psi\psi_L+\mathrm{h.c.}\right)
-|f|^2 -\frac12 D^2
\nn\w2
&&
-\frac{1}{2}D_\mu z D^\mu z^*
-\frac{1}{2}\bar\psi\gamma^\mu D_\mu\psi_L
+2i[z,\bar\psi]\lambda_R
-2i\bar\lambda[\psi_L,z^*]
\Big]\ ,
\label{ymh}
\end{eqnarray}
where
\begin{eqnarray}\label{eq:feqn}
f^*=mz+zz-\frac{1}{n+1}\mathrm{Tr}(zz)\ ,\qquad  D=\frac12[z,z^*]\ ,
\end{eqnarray}
and all fields are in the adjoint representation of $SU(n+1)$. The supersymmetry transformations are given by
\begin{eqnarray}
\delta A_\mu&=&-{\bar\epsilon}\gamma_\mu\lambda\ ,
\nonumber\\
\delta\lambda&=&\frac{1}{8}F_{\mu\nu}\,\gamma^{\mu\nu}\epsilon+\frac{i}{2}D\,\gamma_5\epsilon
\nonumber\\
\delta z&=&\frac{1}{2}\bar\epsilon\psi_L
\nonumber\\
\delta\psi_L&=&D_\mu z\,\gamma^\mu\epsilon_R+f\,\epsilon_L
\label{eq:scalarsusy}
\end{eqnarray}
where $\psi_L=\frac12(1+\gamma_5)\psi$. The conditions for a supersymmetric vacuum are the vanishing of the $f$ and $D$-terms.
We can write a field in the adjoint representation as $z=z_i{}^j$ where $i,j=0,\ldots,n$ are indices of the (anti-)fundamental representation of $SU(n+1)$. With this notation we will consider the supersymmetric vacuum solution given by
\begin{equation}
\langle z_i{}^j\rangle=\frac{1}{n-1}m\delta_i^j-\frac{n+1}{n-1}m\delta_i^0\delta^j_0\,,
\end{equation}
which corresponds to the breaking $SU(n+1)\rightarrow SU(n)\times U(1)$. Next we expand the scalar field around this vacuum as
\begin{equation}
z_i{}^j=\langle z_i{}^j\rangle+\hat z_i{}^j\,.
\end{equation}
Under the symmetry breaking the adjoint scalar $z_i{}^j$ splits up as $z_i{}^j=(z_0{}^0,z_0{}^b,z_a{}^0,z_a{}^b)$ where $a,b=1,\ldots,n$ is an $SU(n)$ (anti-)fundamental index. An infinitesimal gauge-transformation of the broken components of $(z+z^*)_i{}^j$ is given by
\begin{equation}
\delta(z+z^*)_0{}^a=i\Lambda_0{}^i(z+z^*)_i{}^a-i(z+z^*)_0{}^i\Lambda_i{}^a=\frac{2i(n+1) m}{n-1}\Lambda_0{}^a+\ldots
\end{equation}
and similarly for $(z+z^*)_a{}^0$. This means that the real part of $z_0{}^a$ and $z_a{}^0$ can be gauged away by a suitable gauge-transformation $\Lambda_a{}^0$ and $\Lambda_0{}^a$. Therefore by performing a finite gauge-transformation
\begin{equation}
U(\xi)=e^{i\xi^aT_a{}^0+i\xi_aT_0{}^a}\,,
\end{equation}
where $T_0{}^a$ and $T_a{}^0$ are the broken generators which don't preserve the vacuum, we can bring the scalar to the form
\begin{equation}
z_i{}^j=
\left(
\begin{array}{cc}
\frac{-n}{n-1}m +\hat z&i\phi^b\\
i\phi_a&\frac{1}{n-1}m\delta_a^b-\frac{1}{n}\delta_a^b\hat z+\hat z_a{}^b	
\end{array}
\right)\,.
\label{eq:zgauge}
\end{equation}
Note that the scalars $\phi_a$ are now real, in the sense that $(\phi_a)^*=\phi^a$, the imaginary part being eaten by the off-diagonal vectors which become massive. The off-diagonal part of the $SU(n+1)$ gauge field is massive and denoted by
\be
C_\mu^a = (A_\mu)_0{}^a\ ,\qquad C_{\mu a}= (A_\mu)_a{}^0\ .
\ee
Using this notation and \eq{eq:zgauge} in \eq{ymh}, the resulting bosonic Lagrangian is
\bea
{\mathcal L}_{\rm bos} &=&-\frac14 \left( (F^{\mu\nu})_a{}^b + 2iC^{[\mu}_a C^{\nu]^b}\right)
\Big( (F_{\mu\nu})_b{}^a + 2iC_{\mu b}C_\nu^a\Big)-\frac12 G^{\mu\nu}_a G_{\mu\nu}^a
\nn\w2
&&{}
-\frac14 \left( F^{\mu\nu} + 2iC^{[\mu a} C^{\nu]}_a \right)
\left( F_{\mu\nu} + 2iC_\mu^b C_{\nu b}\right) -|f|^2 -\frac12 D^2
\nn\w2
&&{}
-\frac{1}{2}\left|\partial_\mu\hat z-C_\mu^a\phi_a+C_{\mu\,a}\phi^a\right|^2
-\frac{1}{2}\left|D_\mu\hat z_a{}^b-\frac{1}{n}\delta_a^b\partial_\mu\hat z-C_{\mu\,a}\phi^b+C_\mu^b\phi_a\right|^2
\nn\w2
&&{}
-\frac{1}{2}\left|D_\mu\phi^a+\frac{n+1}{n-1}mC_\mu^a-\frac{n+1}{n}\hat zC_\mu^a+C_\mu^b\hat z_b{}^a\right|^2
\nn\w2
&&{}
-\frac{1}{2}\left|D_\mu\phi^a-\frac{n+1}{n-1}mC_\mu^a+\frac{n+1}{n}\hat z^*C_\mu^a+C_\mu^b\hat z^*_b{}^a\right|^2\ ,
\label{ymhbos}
\eea
where $F_{\mu\nu a}{}^b$ and $F_{\mu\nu}$ are the standard $SU(n)$ and $U(1)$ field strengths, while
\be
G_{\mu\nu}^a = 2D_{[\mu}C_{\nu]}^a=2\partial_{[\mu}C_{\nu]}^a+2iA_{[\mu}C_{\nu]}^a+2iC_{[\mu}^c(A_{\nu]})_c{}^a\ .
\ee
is the field strength of the massive vectors. 

In order to make the comparison to the model of massive vectors considered in the previous section we need to truncate the extra scalar multiplets. These consist of the singlet $\hat z$ and the $SU(n)$ adjoint $\hat z_a{}^b$. To see if this truncation is consistent with supersymmetry we should look at the corresponding supersymmetry transformations coming from (\ref{eq:scalarsusy}). These supersymmetry transformations have to be accompanied by a compensating gauge-transformation with (we put $\psi_0{}^a \equiv \chi^a$ and $\psi_a{}^0 \equiv \chi_a$)
\begin{eqnarray}
\Lambda_a{}^0=-\frac{i}{2}R_a{}^b\,\bar\epsilon\chi_b
\qquad
\Lambda_0{}^a=\frac{i}{2}\bar\epsilon\chi^b\,R_b{}^a
\label{eq:compensating}
\end{eqnarray}
where
\begin{equation}
R_a{}^b=
\left(
\frac{2(n+1)}{n-1}\,m\delta_b^a
-\frac{n+1}{n}(\hat z+\hat z^*)\delta_b^a
+\hat z_b{}^a
+\hat z^*_b{}^a
\right)^{-1}
\end{equation}
in order to preserve the gauge (\ref{eq:zgauge}). The full supersymmetry transformations for the Yang-Mills multiplet $(A_{\mu a}{}^b,\lambda_a{}^b)$, the massive Wess-Zumino multiplets $(\hat z_a{}^b, \psi_a{}^b)$ and $(\hat z,\psi)$, and the massive vector multiplet $(C_\mu^a, \phi^a, \chi^a, \lambda^a)$ can be read off from \eq{eq:scalarsusy} by using \eq{eq:zgauge} and the compensating gauge transformations of the fields with gauge parameter given by \eq{eq:compensating}. In particular, the supersymmetry transformations of the scalar fields take the form
\begin{eqnarray}
\delta\hat z&=&\frac{1}{2}\bar\epsilon\psi_L
-\frac{i}{2}\phi^a\,R_a{}^b\,\bar\epsilon\chi_b
-\frac{i}{2}\bar\epsilon\chi^b\,R_b{}^a\,\phi_a
\nonumber\\
\delta\hat z_a{}^b&=&\frac{1}{2}\bar\epsilon\psi_{La}{}^b
+\frac{i}{2}R_a{}^c\,\bar\epsilon\chi_c\,\phi^b
+\frac{i}{2}\phi_a\,\bar\epsilon\chi^c\,R_c{}^b
-\mathrm{trace}\ .
\end{eqnarray}
In a truncation where the scalar fields $(\hat z,{\hat z}_a{}^b)$ are set to zero, we have $R_a{}^b\sim\delta_a^b$, and consequently the vanishing of the variations $\delta\hat z$ and $\delta{\hat z}_a{}^b$  requires that $\chi_a\,\phi^b+\phi_a\,\chi^b=0$. This represents an unacceptable constraint on the fields of the massive vector multiplet. Therefore we conclude that it is not possible (except in the abelian case) to truncate out the additional scalars of the Higgs model and therefore to compare it to the model considered in this paper. In the abelian case the comparison is trivial since all the complicated non-linear terms disappear. The model obtained in this paper by dualizing a massive tensor therefore gives a very different description of massive vectors coupled to super Yang-Mills than that obtained through the Higgs mechanism.

Comparing the bosonic Lagrangians \eq{Lbos} with \eq{ymhbos} shows another key difference. As a test of the tree-level unitarity, considering the process $CC\to CC$ scattering, we see that the model based on \eq{ymhbos} furnishes the vertices $C^2A$, $CCz$ and $C^4$, all of which contribute at the tree level, giving a unitary result in the sense that upon the substitution $C_\mu \to k_\mu/m$ for the longitudinal components of the incoming and outgoing massive vector bosons, all the terms that grow with energy cancel each other out \cite{Bell:1973ex,LlewellynSmith:1973ey,Cornwall:1974km,Horejsi:1993hz}. In the model based on \eq{Lbos}, however, there is only the $C^2 A$ vertex contributing to the same tree-level amplitude which then grows with energy for the longitudinal components. Whether this problem with unitarity can be cured by further extension of our model, or its interpretation as an effective field theory, remains to be investigated.

Motivated by a bosonic Higgsless model where this issue has been studied \cite{Lahiri:2011ic}, in Appendix B we extend the four parameter supersymmetric tensor-Yang-Mills model by introducing two new off-shell invariants, which are possible if we take the 2-form potential in the adjoint representation.


\section{Conclusions}


In coupling $n_V$ vector multiplets to $n_T$ tensor multiplets such that gauge symmetry based on a semi-simple group is realized and that the tensor fields transform in a nontrivial representation of the gauge group, we have seen that the vector fields split into two sets; one belonging to the adjoint representation of the gauge group, and the other that is eaten by as many tensor fields which become massive. Thus we end up with off-shell supersymmetric coupling of a Yang-Mills multiplet to off-shell massive tensor multiplet carrying a nontrivial representation of the gauge group. The massive tensor multiplet also contains scalar fields which are in the same representation. A noteworthy feature of the resulting model is that the elimination of the auxiliary fields needed for the off-shell closure of the algebra gives rise to nonpolynomial couplings of scalar fields to the fermionic bilinears even though the kinetic terms for the scalars is an ordinary one not involving any curved scalar manifold.

Dualization of the massive tensor to a massive vector produces a model in which an on-shell massive nonabelian vector multiplet is coupled to an off-shell Yang-Mills multiplet. The two arbitrary parameters of the model are the mass $M$ and the angle $\theta$. The $\theta \to 0$ limit exists but the $M\to 0$ limit is singular. In the abelian limit the $M\to 0$ limit exists but the resulting Lagrangian is not supersymmetric.

The bosonic sector of the nonabelian model differs from a bosonic Yang-Mills-Higgs system in which the Yang-Mills group $G$ is broken down to $H$ with massive vector fields in representations of $H$, and presence of scalar multiplets. In the supersymmetric Proca-Yang-Mills system, however, we start with a Yang-Mills group H from the beginning and couple it to massive vector multiplets  only without introducing any scalar multiplets. We showed that a consistent supersymmetric truncation of the scalar multiplets in the Yang-Mills-Higgs system is not possible, showing that the two models are genuinely distinct.
In Section 6, we discussed the issue of breakdown in tree-level unitarity in the Proca-Yang-Mills system, which is of course absent in the Yang-Mills-Higgs system. In Appendix B, we presented a 2-parameter extension of the model which has implications for solving this problem though the full study of this extended model remains to be done.

Possible generalizations of the massive tensor-Yang-Mills system presented here are as follows. The tensor hierarchy can be extended to include the 3-form and 4-form potentials. Non-semisimple groups may be considered for completeness. Coupling to scalar multiplets would generalize the results of \cite{Louis:2004xi} where such couplings in the case of the abelian model have been given. Coupling of the nonabelian model to supergravity can be carried out and that would generalize the result of \cite{Mukhi:1979wc} where a single massive vector multiplet, and \cite{Nishino:2012sm} where a truncated version of our full non-abelian model have been coupled to supergravity. Finally, the computation of the full  tree-level amplitude for the scattering of massive vectors in the extended model given in Appendix B, and its implications for unitarity is an interesting open problem.

\section*{Acknowledgements}

We thank B. Dutta, Tianjun Li, W. Schulgin, H. Samtleben and R. Wimmer for useful discussions. This research is supported in part by NSF grant PHY-0906222.

\newpage

\appendix


\section{Dimensional reduction from $D=6$}


The supersymmetry transformations of the $D=4$ $N=1$ tensor-YM system considered in this paper can be obtained from the corresponding $N=(1,0)$ supersymmetric system in $D=6$ \cite{Samtleben:2012mi}:
\begin{eqnarray}
\delta \phi^a &=& \bar\epsilon\chi^a\,,\nonumber\\
\delta {\mathcal B}_{\mu\nu}^a &=& 2 \, \bar\epsilon\,\gamma_{[\mu}D_{\nu]} \lambda^a -\bar\epsilon\gamma_{\mu\nu}\chi^a\,,
\nonumber\\
\delta \lambda^{i\,a} &=& \frac{1}{8}\,\gamma^{\mu\nu}\mathcal B^a_{\mu\nu} \epsilon^i
-\frac{1}{2}\,Y^{ij\,a}\epsilon_j + \frac14  \phi^a \epsilon^i \,,
\nonumber\\
\delta\chi^{i\,a} &=&\frac{1}{48}\gamma^{\mu\nu\rho}\,\mathcal H_{\mu\nu\rho}^a\,\epsilon^i
+\frac{1}{4}\,\gamma^\mu D_\mu\phi^a \epsilon^i
-\frac{1}{2}(T_{\alpha})_b{}^a\, \gamma^\mu\lambda^{i\,(\alpha}\, \bar\epsilon\gamma_\mu \lambda^{b)}\,,
\nonumber\\
\delta Y^{ij\,a}&=&-{\bar\epsilon}^{(i}\gamma^\mu D_\mu\lambda^{j)a}+2\,\bar\epsilon^{(i} \chi^{j) a}\,.
\end{eqnarray}
where $i=1,2$ and the spinors are symplectic-Majorana-Weyl. The $D=4$ system is obtained by making the following ansatz
\begin{eqnarray}
\sigma_2\epsilon &=& \gamma_5 \epsilon\ ,\quad \tau_2\epsilon=\gamma_5\epsilon\,,
\quad \sigma_2\lambda = \gamma_5 \lambda\ ,\quad \tau_2\lambda=\gamma_5\lambda\,,
\quad \sigma_2\chi = \gamma_5 \chi\ ,\quad \tau_2\chi=-\gamma_5\chi\,,
\nonumber\\
A_4&=&0\ , \quad A_5=0\ ,\quad {\mathcal B}_{\mu 4}=0\ ,\quad {\mathcal B}_{\mu 5}=0\,,
\quad {\mathcal B}_{45} \equiv \varphi\ , \quad Y_i{}^j= (i\tau_2)_i{}^j\,D\,,
\end{eqnarray}
where we have suppressed the representation indices, $D$ is a real scalar, the Pauli $\tau$-matrices act on the R-symmetry indices and the Pauli $\sigma$-matrices act on the $SO(2)$ doublet indices where $SO(2)$ is contained in $SO(5,1)\sim SO(3,1)\times SO(2)$. We also use the convention by which the 6D Dirac $\Gamma$-matrices are $\Gamma^\mu=\gamma^\mu\times 1\ , \Gamma^4=\gamma^5\times \sigma^3$ and $\Gamma^5=\gamma_5\times\sigma^1$. The resulting $N=1, D=4$ supersymmetry transformation rules are
\begin{eqnarray}
\delta\phi^a&=&\bar\epsilon\chi^a\,,
\nonumber\\
\delta\varphi^a&=&i\bar\epsilon\gamma_5\chi^a\,,
\nonumber\\
\delta\mathcal B_{\mu\nu}^a &=&2\bar\epsilon\,\gamma_{[\mu}D_{\nu]}\lambda^a
-\bar\epsilon\gamma_{\mu\nu}\chi^a\,,
\nonumber\\
\delta\chi^a&=&
\frac{1}{48}\mathcal H_{\mu\nu\rho}^a\,\gamma^{\mu\nu\rho}\epsilon
+\frac{i}{8}D_\mu\varphi^a\,\gamma^\mu\gamma_5\epsilon
+\frac{1}{4}D_\mu\phi^a\,\gamma^\mu\epsilon
-\frac{1}{2}(T_\alpha)_b{}^a\,\gamma^\mu\lambda^{(\alpha}\,\bar\epsilon\gamma_\mu\lambda^{b)}\,,
\nonumber\\
\delta \lambda^a &=& \frac{1}{8}\mathcal B^a_{\mu\nu}\,\gamma^{\mu\nu}\epsilon
+\frac{i}{4}\varphi^a\,\gamma^5\epsilon
+\frac{i}{2}D^a\,\gamma_5\epsilon
+\frac{1}{4}\phi^a\,\epsilon\,,
\nonumber\\
\delta D^a&=&\frac{i}{2}\bar\epsilon \gamma_5\gamma^\mu D_\mu\lambda^a
-i\bar\epsilon\gamma_5\chi^a\,.
\end{eqnarray}
This is not yet the off-shell system that we want since there is an extra scalar $\varphi^a$. It turns out that the closure of this algebra requires, among other things, that
\begin{equation}
D^\mu\varphi^a
=
\frac{1}{6}\varepsilon^{\mu\nu\rho\sigma}\mathcal H^a_{\nu\rho\sigma}
-2i(T_\alpha)_b{}^a\,{\bar\lambda}^b\gamma^\mu\gamma_5\lambda^\alpha\ .
\end{equation}
Using this fact in the expression for $\delta\chi^a$ and performing the shift $D^a\rightarrow D^a-\frac{1}{2}\varphi^a$ we obtain the algebra
\begin{eqnarray}
\delta\phi^a&=&\bar\epsilon\chi^a\,,
\nonumber\\
\delta\mathcal B_{\mu\nu}^a &=&2\bar\epsilon\,\gamma_{[\mu}D_{\nu]}\lambda^a
-\bar\epsilon\gamma_{\mu\nu}\chi^a\,,
\nonumber\\
\delta\chi^a&=&
\frac{1}{24}\mathcal H_{\mu\nu\rho}^a\,\gamma^{\mu\nu\rho}\epsilon
+\frac{1}{4}D_\mu\phi^a\,\gamma^\mu\epsilon
-\frac{1}{2}(T_\alpha)_b{}^a\,(\gamma^\mu\lambda^{(\alpha})\,\bar\epsilon\gamma_\mu\lambda^{b)}\,,
+\frac{1}{4}(T_\alpha)_b{}^a\,(\gamma^\mu\gamma_5\epsilon)\,{\bar\lambda}^b\gamma_\mu\gamma_5\lambda^\alpha
\nonumber\\
\delta\lambda^a&=&\frac{1}{8}\mathcal B^a_{\mu\nu}\,\gamma^{\mu\nu}\epsilon
+\frac{i}{2}D^a\,\gamma_5\epsilon
+\frac{1}{4}\phi^a\,\epsilon\,,
\nonumber\\
\delta D^a&=&\frac{i}{2}\bar\epsilon \gamma_5\gamma^\mu D_\mu\lambda^a
-\frac{i}{2}\bar\epsilon\gamma_5\chi^a\,.
\end{eqnarray}
This coincides with the algebra in (\ref{eq:tensorYM}).


\section{An extension of the model}


In Section 6, we compared our Higgsless model based on \eq{master} for the description of supersymmetric Proca-Yang-Mills system with the supersymmetric Yang-Mills-Higgs system and we noted the presence of divergences in the tree level amplitude for the scattering of the longitudinally polarized massive vectors in the Higgsless model. One attempt in constructing a Higgsless model that avoids this problem is based on the Lagrangian \cite{Lahiri:2011ic}
\be
{\mathcal L}= {\rm Tr}\, ( -\frac14 F^2 +\frac1{12}H^2+  m B\wedge F )\ ,
\label{badL}
\ee
where $F=dA+A^2$ and the 2-form potential has the field strength $H=dB+ [A,B]$. Diagonalization of the kinetic terms gives a massive vector boson and the potential divergences in the scattering of two such longitudinally polarized vectors turn out to cancel out \cite{Lahiri:2011ic}. An important shortcoming of this model not addressed in \cite{Lahiri:2011ic} is that the divergence of the the field equation for the 2-form potential gives an unacceptable constraint; see the discussion around (3.6) above, where this type constraint is avoided by introducing an explicit mass term for the 2-form potential. Nonetheless, given that new couplings relevant for the unitarity issue arise due the $B\wedge F$ term in (\ref{badL}), here we shall extend our supersymmetric Proca-Yang-Mills model by considering the case when the massive supermultiplet is taken to be in the adjoint representation of the gauge group. In this case, the following two invariants can be added to the action (\ref{master}) (where needed we use subscripts $1,2$ to distinguish fields of the Yang-Mills and tensor multiplet respectively)
\bea
\mathcal L_{extra} &=&
\alpha m\, {\rm Tr} \left( \mathcal B_{\mu\nu}F^{\mu\nu}
-8\bar\lambda_1\chi +8\bar\lambda_1{\slashed D}\lambda_2 -8D_1D_2\right)
\nn\w2
&& +\beta m\,{\rm Tr} \left(\varepsilon^{\mu\nu\rho\sigma}\mathcal B_{\mu\nu}F_{\rho\sigma}
-16i\bar\lambda_1\gamma_5\chi +8\phi\,D_1\right)\ .
\eea
Adding this Lagrangian to that given in \eq{master} gives the total bosonic Lagrangian
\begin{eqnarray}
{\mathcal L''}_{\mathrm{bos}}&=& {\rm Tr} \Bigl(-\frac{1}{4}F_{\mu\nu}F^{\mu\nu}
-\frac{1}{6}\mathcal H_{\mu\nu\rho}\mathcal H^{\mu\nu\rho}
-\frac12 \mathcal B_{\mu\nu}
\left( m^2 \mathcal B^{\mu\nu}+\theta m^2*\mathcal B^{\mu\nu}-2m\alpha F^{\mu\nu} -4m\beta*F^{\mu\nu}\right)
\nn\w2
&&{}
-D_\mu \phi D^\mu\phi-m^2 \phi^2
+4m\phi\left(2\beta D_1+\theta m D_2\right)
+2\left(D_1^2-4m\alpha D_1D_2 +2 m^2D_2^2\right)
\nn\w2
&& -\phi[F_{\mu\nu},{\mathcal B}^{\mu\nu}]+8\phi[D_1,D_2] \Bigr) \ .
\end{eqnarray}
Integrating out the auxiliary fields gives, modulo the fermions,
\begin{eqnarray}
D_1 &=& 2m(\alpha D_2 -\beta \phi)+2[\phi,D_2]
\label{d1}\\
m^2D_2 &=&-\frac12 \theta m^2 \phi-[\phi,D_1]+m\alpha D_1\ .
\label{d2}
\end{eqnarray}
Solving for $D_1^\alpha = 2 {\rm Tr}\,(T^\alpha D_1) $, up to fermions, gives
\begin{eqnarray}
D_1^\alpha
&=&
-m \left(\theta\alpha+2\beta\right) \left(\delta_{\alpha\beta}-2\alpha^2\delta_{\alpha\beta}
+\frac{2}{m^2}f_{\alpha\gamma}{}^\delta f_{\beta\delta}{}^\kappa\,\phi^\gamma\phi_\kappa\,\right)^{-1} \phi^\beta\ ,
\label{d2sol}
\end{eqnarray}
and substitution into \eq{d2} gives $D_2$. Note that, unlike in the case of the model given in section 3, here the elimination of the auxiliary fields generate a potential. 

Upon dualizing the massive 2-form potential to a massive vector field, the resulting bosonic terms in the Lagrangian take the form
\begin{eqnarray}
\mathcal L'_{\mathrm{bos}}&=& {\rm Tr}\,\Big[
-\frac{1}{4}F_{\mu\nu}F^{\mu\nu}
+\frac{m^2}{2M^2}(\alpha^2+4\theta\alpha\beta-4\beta^2)F_{\mu\nu} F^{\mu\nu}
-M^2 C_\mu C^\mu
\nonumber\\
&&{}
-\frac{m^2}{2M^2}(\theta \alpha^2-4\alpha\beta-4\theta\beta^2)F_{\mu\nu}*F^{\mu\nu}
-\frac{1}{2}G^{\mu\nu}\left(G_{\mu\nu}-\theta *G^{\mu\nu}\right)
\nonumber\\
&&{}
+\frac{m}{M}F^{\mu\nu}\Big[(\theta\alpha-2\beta)G_{\mu\nu}
+(\alpha+2\theta\beta)*G_{\mu\nu}\Big]
-D_\mu\phi D^\mu\phi-m^2\phi^2
\nonumber\\
&&{}
+2D_1^2+8m\beta\,\phi\,D_1-8m\alpha\,D_1\,D_2+4\theta m^2\phi D_2+4m^2D_2^2
\nonumber\\
&&{}
+\frac1M\phi\,[*\, G_{\mu\nu},F^{\mu\nu}-\theta * F^{\mu\nu}]+8\phi [D_1,D_2]\,\Big]
\nonumber\\
&&{}
-\frac{1}{4M^2} f_{\alpha\gamma}{}^\kappa f_{\beta\kappa}{}^\delta\,\phi_\delta\phi^\gamma F^{\mu\nu\,\alpha}(F^\beta_{\mu\nu}-\theta*F^\beta_{\mu\nu})\ .
\end{eqnarray}
Substituting for $D_1$ and $D_2$ the expressions given in \eq{d2sol} and \eq{d2} yields a potential that vanishes in the absence of the parameters $\alpha$ and $\beta$. Considering the tree level amplitude for the $CC\to CC$ scattering, we observe that, in addition to the $C^2 A$ vertex which contributes to the tree graph, the introduction of the $(\alpha,\beta)$ parameters gives rise to a mixing $AC$ which together with the usual $A^3$ and $A^4$ couplings, gives rise to new contributions. However, a detailed computation and analysis of the tree level $CC\to CC$ scattering amplitude remains to be performed.

\newpage

\bibliography{Master}{}

\bibliographystyle{utphys}

\end{document}